\documentstyle[aps,prl,multicol,amsfonts]{revtex}
\tighten
\draft
\newtheorem{lemma}{Lemma}
\newtheorem{claim}{Claim}
\newtheorem{define}{Definition}
\newtheorem{theorem}{Theorem}
\begin{document}
\title{One Dimensional $n$ary Density Classification Using Two Cellular
 Automaton Rules}
\author{H. F. Chau\footnote{electronic address: hfchau@hkusua.hku.hk}, L. W.
 Siu and K. K. Yan}
\address{Department of Physics, University of Hong Kong, Pokfulam Road, Hong
 Kong}
\date{\today}
\maketitle
\begin{abstract}
 Suppose each site on a one-dimensional chain with periodic boundary condition
 may take on any one of the states $0,1,\ldots ,n\!-\!1$, can you find out
 the most frequently occurring state using cellular automaton? Here, we prove
 that while the above density classification task cannot be resolved by a
 single cellular automaton, this task can be performed efficiently by applying
 two cellular automaton rules in succession.
\end{abstract}
\medskip
\begin{multicols}{2}
 Cellular automaton (CA) is a simple local parallel interaction model for
 many natural systems \cite{Wolfram,Def}. And from the computer science point
 of view, CA can be regarded as a special kind of Turing machine without
 internal memory. In fact, tailor-made CA can be used to simulate certain
 logical operations \cite{CA_Appl}.
\par
 Since CA is essentially an internal-memoryless Turing machine, it is natural
 to ask if it can be used to perform certain tasks that require global
 counters. One such task, called (binary) density classification, was recently
 studied by Land and Belew \cite{Land_Belew}. They consider a one-dimensional
 (finite but arbitrarily long) chain of sites with periodic boundary condition.
 Each site is in either state zero or state one. That is, each site contains a
 Boolean state. Our task is to change the state of every site to one if the
 number of ones is more than the number of zeros in the chain. (That is, the
 density of one, $\rho_c$, in the chain is greater than 1/2.) Otherwise, every
 site is set to the state zero. Clearly, the density classification problem is
 trivial if one uses a global counter. Alternatively, one can also solve this
 problem if the CA rule table scales with the number of sites $N$ so that the
 CA model becomes nonlocal in the limit of large $N$. Nonetheless, Land and
 Belew proved that (binary) density classification cannot be done perfectly
 using a single one-dimensional CA \cite{Land_Belew}. Their proof can be
 extended to multiple dimensions, too.
\par
 Quite unexpectedly, Fuk\'{s} found that the (binary) density classification
 problem can be solved if we apply {\em two} CA rules in succession
 \cite{Fuks}. Later on, Chau {\em et al.} generalized his result by showing
 that classifying any rational density $\rho_c$ on a $N$ site chain can be
 performed efficiently in $\mbox{O}(N)$ time using two CA rules in succession
 \cite{Chau}.
\par
 At this point, it is natural to ask if CA can be used to classify $n$ary
 density, namely, when each site takes on a $n$ary state. In fact, Land and
 Belew conjectured that a significantly different way of argument is required
 to generalize their ``no-go'' result to the $n$ary case \cite{Land_Belew}. In
 this Letter, we first prove that $n$ary density classification by CA is
 impossible. Although we only report our proof for the one-dimensional case,
 our arguments can be readily generalized to multiple dimensions. But most
 important of all, our proof provides a hint to solve the $n$ary density
 classification problem using two CA rules. Guided by this hint, we report a
 simple and efficient way to classify $n$ary density with two CA rules.
\par
 {\em Statement of the problem} --- We begin by formally defining the $n$ary
 density classification problem. Suppose that each of the one-dimensional chain
 of $N$ sites in periodic boundary condition may take on a state in $0,1,\ldots
 ,n\!-\!1$. We define $\rho_i (\alpha)$ as the number of sites in state $i$
 for the configuration $\alpha$ divided by $N$. That is, $\rho_i (\alpha)$ is
 the density of the state $i$ of the configuration $\alpha$. Our goal is to
 evolve the state of all sites to $i$ if and only if $\rho_i (\alpha) > \rho_j
 (\alpha)$ for all $j\neq i$. While this task is trivial if one has a global
 counter, we now show that this task is not achievable using a (deterministic)
 CA. More precisely, we mean that given any deterministic CA with a finite rule
 table, we can find a configuration with a sufficiently large $N$ such that
 this configuration cannot be correctly classified by this CA.
\par
 {\em Impossibility to classify density by one CA rule} --- We prove our
 impossibility result by contradiction. Suppose a CA rule with radius $r$
 correctly classifies any $n$ary density for all $N$, then we denote the action
 of this CA on a configuration $\alpha$ by ${\mathbf T}$. Moreover, we denote
 the set of all configurations with density $\rho_i > \rho_j$ for all $j\neq i$
 by $\Omega_i$. Then, Land and Belew showed that \cite{Land_Belew}
\begin{lemma} ${\mathbf T} \left[\Omega_i \right] \subset \Omega_i$ for all
 $i$. Besides, ${\mathbf T} \left[ j^N \right] = j^N$ where $j^N$ denotes a
 configuration of $N$ consecutive $j$s. In fact, if the states of a site and
 its $2r$ neighbors are all $j$, then the state of that site under the action
 of ${\mathbf T}$ equals $j$. Similarly, if $\alpha$ is a period $2r\!+\!1$
 sequence, then so is ${\mathbf T} (\alpha)$. \label{Lemma:Invariant}
\end{lemma}
\indent
 Now, we make the following claim:
\begin{claim}
 The action of ${\mathbf T}$ on any configuration $\alpha$ preserves the
 densities $\rho_i$ of the configuration $\alpha$. \label{Claim:Invariant}
\end{claim}
\par\indent
 Clearly, our conclusion that no such CA rule ${\mathbf T}$ exits follows
 directly from Claim~\ref{Claim:Invariant} because the state $00\ldots 001$
 cannot be brought to $00\ldots 000$ under the repeated action of
 ${\mathbf T}$.
\par\medskip\noindent
{\em Proof of the claim:} Suppose the contrary, then we can find a
 configuration $\beta$ whose density $\rho_i$ is not preserved under the action
 of ${\mathbf T}$ for some $i$. Without lost of generality, we may assume that
 $\rho_0 (\beta) > \rho_0 ({\mathbf T}(\beta))$ and $\rho_1 (\beta) < \rho_1
 ({\mathbf T}(\beta))$. Otherwise the proof is similar. If we denote the length
 of $\beta$ by $s$, then $\rho_0 ({\mathbf T}(\beta)) \leq \rho_0 (\beta) -
 1/s$. From Lemma~\ref{Lemma:Invariant}, we know that $\beta$ must make up of
 more than one type of states. Now, we consider configurations in the form
 $\gamma = \beta^x 0^y 1^z$; that is, $\gamma$ makes up of $x$ copies of
 $\beta$ followed by $y$ consecutive zeros and then $z$ consecutive ones. We
 choose the ratio $x\!:\!y\!:\!z$ in such a way that $\gamma\in\Omega_0$ while
 $\rho_1 (\gamma) < \rho_0 (\gamma) < \rho_1 (\gamma) + \frac{x}{2(s x+y+z)}$.
 In other words, $\gamma$ has a slight excess of zeros over ones to put it in
 $\Omega_0$. Now, we consider the state ${\mathbf T}(\gamma)$. Since the radius
 of our CA equals $r$, so except for the $6r$ sites at the boundaries between
 $\beta$, $0$ and $1$, Lemma~\ref{Lemma:Invariant} tells us that states of all
 other sites containing $0^y 1^z$ are unchanged by the action of ${\mathbf T}$.
 Consequently, $\rho_0 ({\mathbf T}(\gamma)) \leq \rho_0 (\gamma) - \frac{x}{s
 x+y+z} + \frac{4r}{s x+y+z}$. So, if we choose sufficiently large values of
 $x$, $y$ and $z$ while keeping the ratio $x\!:\!y\!:\!z$ fixed, we can make
 $\rho_0 ({\mathbf T}(\gamma)) \leq \rho_0 (\gamma) - \frac{x}{2(s x+y+z)}$
 while $\rho_1 ({\mathbf T}(\gamma)) > \rho_1 (\gamma) + \frac{x}{2(s x+y+z)}$.
 Therefore, $\rho_0 ({\mathbf T}(\gamma)) < \rho_1 ({\mathbf T}(\gamma))$.
 Hence ${\mathbf T} (\gamma) \not\in \Omega_0$ although $\gamma\in\Omega_0$.
 This contradicts Lemma~\ref{Lemma:Invariant}.
\hfill$\Box$
\par\medskip\indent
 Clearly, we can modify the above proof to show that $n$ary density
 classification in multiple dimension by a single CA is also impossible. More
 importantly, Claim~\ref{Claim:Invariant} tells us that in order to construct
 an $n$ary density classification rule using two CAs, the first rule must
 preserve densities. This is the reason why both the traffic rule and modified
 traffic rule used by Fuk\'{s} \cite{Fuks} and Chau {\em et al.} \cite{Chau} in
 binary density classification are density preserving.
\par
 {\em Classification of $n$ary density using two CA rules} --- Let us introduce
 a few useful definitions before reporting the two CA rules that classifies
 $n$ary density.
\begin{define}
 Let $\alpha = (a_i)_{i=1}^N$ be a one-dimensional configuration in periodic
 boundary condition. An elementary block (EB) in $\alpha$ is defined to be a
 subsequence of consecutive states $(b_j)_{j=1}^k$ of $\alpha$ with $b_1 < b_2
 < \cdots < b_k$. Moreover, such a subsequence is maximal in the sense that
 inclusion of further element into the subsequence in either ends does not
 produce any EB. \label{Def:Ele_Block}
\end{define}
\par\indent
 Clearly, any configuration $\alpha$ can be uniquely decomposed into a
 collection of EBs. For instance, the configuration $40223441$ makes up of EBs
 $02$, $234$, $4$ and $14$. Furthermore, for an $n$ary state system, elementary
 combinatorics arguments shows that there are totally $2^n\!-\!1$ different
 possible EBs.
\begin{define}
 We define the homogeneous lexicographic (hlex) ordering \cite{Order} to these
 $2^n\!-\!1$ EBs as follows: An EB $(a_j)$ is greater than another EB $(b_j)$
 if and only if (1) the length of $(a_j)$ is greater than that of $(b_j)$; or
 (2) the lengths of $(a_j)$ and $(b_j)$ agree and $a_j > b_j$ for the
 {\em first} index $j$ with $a_j \neq b_j$. \label{Def:Order}
\end{define}
\par\indent
 We denote the $2^n\!-\!1$ EBs by ${\cal B}_i$s. Then, the hlex order goes as
 ${\cal B}_0 \equiv 0 <_{\rm hlex} {\cal B}_1 \equiv 1 <_{\rm hlex} \cdots
 <_{\rm hlex} {\cal B}_{n-1} \equiv n\!-\!1 <_{\rm hlex} {\cal B}_n \equiv 01
 <_{\rm hlex} {\cal B}_{n+1} \equiv 02 \cdots <_{\rm hlex} {\cal B}_{2n-2}
 \equiv 0(n\!-\!1) <_{\rm hlex} \cdots <_{\rm hlex} {\cal B}_{2^n-2} \equiv
 012\ldots (n\!-\!1)$. We may interpret the hlex ordering as a measure of
 affinity between different states.
\par
 Our goal of the first CA rule is to make the EBs to interact with each other,
 hoping to make as many high affinity EBs as possible. We achieve this goal by
 first introducing a {\em greedy interaction} between two adjacent EBs. Since
 the union and intersection of elements of two EBs can form EBs. So, we have:
\begin{define}
 Let ${\cal C}_1 {\cal C}_2$ be two adjacent EBs, and we abuse the notation a
 bit by writing ${\cal C}_1$ and ${\cal C}_2$ both as sequences and sets. Then,
 we define $I({\cal C}_1 {\cal C}_2) = {\cal D}_1 {\cal D}_2$ where
 ${\cal D}_1$ is the EB formed by the set ${\cal C}_1 \cup {\cal C}_2$, and
 ${\cal D}_2$ is the (possibly empty) EB formed by the elements in ${\cal C}_1
 \cap {\cal C}_2$. In case ${\cal D}_1 {\cal D}_2$ equals ${\cal C}_1
 {\cal C}_2$ or ${\cal C}_2 {\cal C}_1$, then we said that ${\cal C}_1$
 interacts elastically with ${\cal C}_2$. Otherwise, we said that ${\cal C}_1$
 interacts inelastically with ${\cal C}_2$. \label{Def:Interaction}
\end{define}
\par\indent
 From the above definition, readers can verify that $I(2301) = 0123$, $I(02123)
 = 01232$, $I(01201) = 01201$ and $I(0101) = 0101$. We also observe that
\begin{lemma}
 The EB ${\cal D}_1$ defined above is the highest possible EB formed from
 ${\cal C}_1 \cup {\cal C}_2$ with respected to the hlex order. Furthermore,
 ${\cal D}_2 \leq_{\rm hlex} {\cal D}_1$. \label{Lemma:Interaction_Ordering}
\end{lemma}
\par\medskip\noindent
{\em Proof:} Since all elements in an EB are distinct, so the highest possible
 affinity EB is formed by selecting all the elements in ${\cal C}_1 \cup
 {\cal C}_2$. In addition, ${\cal D}_2 \leq_{\rm hlex} {\cal D}_1$ follows
 directly from the fact that ${\cal C}_1 \cap {\cal C}_2 \subset {\cal C}_1
 \cup {\cal C}_2$.
\hfill$\Box$
\par\medskip\indent
 Since ${\cal D}_1$ is the highest possible affinity EB constructed out of
 ${\cal C}_1$ and ${\cal C}_2$, the greedy interaction is an effective way to
 construct high affinity while keeping the density $\rho_i$ of a configuration.
 Now, we report two technical lemmas before going on.
\begin{lemma}
 Let $\alpha$ be a one-dimensional configuration with $N$ sites. Suppose there
 exist two EBs in $\alpha$ which can interact inelastically with each other,
 then the total number of EBs in $\alpha$ is less than or equal to $N\!-\!2$.
 Besides, this bound is tight. \label{Lemma:EB_Count}
\end{lemma}
\par\medskip\noindent
{\em Proof:} Suppose $\alpha$ contains $N$ EBs, then from
 Definition~\ref{Def:Ele_Block}, it is easy to see that states in all the $N$
 sites must be identical. Hence, none of the EB interacts inelastically with
 each other. Similarly, if $\alpha$ contains $N\!-\!1$ EBs, then exactly
 $N\!-\!2$ of them contains one element and one of them contains two elements.
 Since periodic boundary conditions applies to the $N$ sites, again from
 Definition~\ref{Def:Ele_Block}, we know that all the $N\!-\!2$ singleton EBs
 must be in an identical state, say $a$. Suppose the remaining EB is in state
 $bc$, then $b\leq a\leq c$. (Otherwise, we can extend the size of the EB $bc$
 by one, contradicting the maximality of an EB.) Now, it is easy to check that
 $a$ and $bc$ interact elastically. Thus, an inelastically interacting
 configuration must contain less than $N\!-\!1$ EBs. Finally, the tightness of
 this bound is revealed by the configuration $020100000$.
\hfill$\Box$
\begin{lemma}
 Let $\alpha$ be a configuration whose EB only interacts elastically with each
 other. Then $\alpha\in\Omega_i$ if and only if ${\cal B}_i$ is an EB in
 $\alpha$. \label{Lemma:Omega_i}
\end{lemma}
\par\medskip\noindent
{\em Proof:} From Definition~\ref{Def:Interaction}, two EBs interacts
 elastically if and only if one is a subset of the other. Hence, if the EBs in
 $\alpha$ interact elastically, we can always arrange them sequentially so that
 one is a subset of the next. Clearly, the most frequently occurring state(s)
 are the ones in the first element of this sequence ${\cal C}$. Thus, $\alpha
 \in\Omega_i$ if and only if $i$ is the only element in ${\cal C}$. Hence, our
 assertion is proved.
\hfill$\Box$
\par\medskip\indent
 At this point, we have introduced enough material to present our first CA rule
 for the $n$ary density classification problem. We express a configuration
 $\alpha$ in terms of its EBs ${\cal C}_1 {\cal C}_2 \cdots {\cal C}_j$.
 Inspired by the traffic rule used in Ref.~\cite{Land_Belew}, we define the
 basic rules ${\mathbf R}_i$ for $i=0,1,\ldots ,2^n\!-\!2$ as follows: Each EB
 ${\cal C}_k \subset \alpha$ with ${\cal C}_k = {\cal B}_i$ will interact with
 its forward neighboring EB ${\cal C}_{k+1}$ if and only if ${\cal C}_{k+1}
 \neq {\cal B}_i$. Besides, all interactions are taken in parallel. For
 instance, ${\mathbf R}_1 (1021101) = 0121011$. Clearly, ${\mathbf R}_i$
 conserves $\rho_j$. More importantly, it is straight-forward to check that
 ${\mathbf R}_i$ is equivalent to the following rules: (1) If ${\cal C}_{k-1}$,
 ${\cal C}_k \neq {\cal B}_i$ or ${\cal C}_k$, ${\cal C}_{k+1} = {\cal B}_i$,
 then the states of the sites occupied by ${\cal C}_k$ remain unchanged; (2)
 Otherwise, if ${\cal C}_{k-1} = {\cal B}_i$, then the states of the sites
 occupied by ${\cal C}_k$ become the last $\ell ({\cal C}_k)$ states of $I
 ({\cal C}_{k-1} {\cal C}_k)$ where $\ell ({\cal C}_k)$ denotes the length of
 the sequence ${\cal C}_k$; (3) For the remaining case that ${\cal C}_{k+1}
 \neq {\cal B}_i$, then the states of the sites occupied by ${\cal C}_k$ become
 the first $\ell ({\cal C}_k)$ states of $I({\cal C}_k {\cal C}_{k+1})$. Hence,
 ${\mathbf R}_i$ is a CA rule with radius $n\!+\!\ell ({\cal B}_i)\!-\!1$.
\par
 Since finite composition of CA rules is also a CA rule, we have:
\begin{define}
 We write the {\em affinity rule} ${\mathbf A} = {\mathbf R}_0 \circ \left(
 {\mathbf R}_0 \circ {\mathbf R}_1 \right)$ $\circ \left( {\mathbf R}_0 \circ
 {\mathbf R}_1 \circ {\mathbf R}_2 \right)$ $\circ \cdots \circ \left(
 {\mathbf R}_0 \circ {\mathbf R}_1 \circ \cdots \circ {\mathbf R}_{2^n-2}
 \right)$. (That is, a total of $(2^{n-1}\!-\!1)(2^n\!-\!3)$ terms in the above
 composition.) Then, the CA rule ${\mathbf A}$ preserves $\rho_i$ and has a
 radius between $n (2^{n-1}-1)(2^n-3)$ and $(2n\!-\!1) (2^{n-1}\!-\!3)
 (2^n\!-\!3)$. \label{Def:Affinity_Rule}
\end{define}
\par\indent
 Note that the mobility of an EB increases with decreasing affinity, so the
 spirit of our affinity rule is to aggressively produce a much high affinity
 EBs as we can. In fact, we find that
\begin{lemma}
 The total number of EBs of a configuration $\alpha$ is greater than or equal
 to that of ${\mathbf A} (\alpha)$. Besides, under the repeated action of
 ${\mathbf A}$, the number of EBs in a configuration will eventually stay
 constant. \label{Lemma:Non_Decreasing}
\end{lemma}
\par\medskip\noindent
{\em Proof:} The first part of the lemma follows directly from
 Definitions~\ref{Def:Interaction} and~\ref{Def:Affinity_Rule}. Since $N$ is
 finite, the total number of EBs for any configuration must lie between 1 and
 $N$. Since the number of EBs is a decreasing function of ${\mathbf A}$ and $n$
 is finite, so under the repeated action of ${\mathbf A}$, the number of EBs in
 a configuration will eventually stay constant.
\hfill$\Box$
\par\medskip\indent
 With all the preparation works above, we have confident to use ${\mathbf A}$
 as our first CA rule in the $n$ary density classification problem. The power
 of this CA rule is apparent from the Theorem below.
\begin{theorem}
 Let $\alpha$ be a configuration in a one-dimensional chain of $N$ sites. Then
 $\alpha\in\Omega_i$ if and only if ${\mathbf A}^{N-2} (\alpha)$ contains an EB
 ${\cal B}_i$. \label{Thrm:Time_Affinity_Rule}
\end{theorem}
\par\medskip\noindent
{\em Proof:} By Lemma~\ref{Lemma:Omega_i}, we only need to prove that for any
 configuration $\alpha$, ${\mathbf A}^{N-3} (\alpha)$ does not contain
 inelastically interacting EBs. In addition, we may assume that $\alpha$
 contains at least a pair of inelastically interacting EBs. Otherwise, our
 assertion is trivially true.
\par
 We follow the motion of an arbitrarily chosen EB under the action repeated of
 ${\mathbf A}$. After the EB interacts with another one, then we turn to follow
 the motion of the left resultant EB. Besides, we turn to follow the motion of
 the left neighboring EB in case the left neighboring EB equals to that of our
 current EB. So, Lemma~\ref{Lemma:Interaction_Ordering} tells us that the
 affinity of the EB we are following increases with time. In addition,
 Definition~\ref{Def:Affinity_Rule} tells us that EBs with different affinity
 interact at a different rate. More importantly, by direct checking, one sees
 that our monitoring EB must interact with the left neighboring EB under the
 action of ${\mathbf A}$. Besides, under the action of ${\mathbf A}$, two
 distance EBs can interact only when one of them first hops through all the
 EBs separating them by elastic interactions. Now, we denote the highest
 affinity EB under our monitoring scheme in ${\mathbf A}^{N-3} (\alpha)$ by
 ${\cal C}_1$, then from Lemmas~\ref{Lemma:Interaction_Ordering},
 \ref{Lemma:EB_Count} and~\ref{Lemma:Non_Decreasing}, all EBs in
 ${\mathbf A}^{N-3} (\alpha)$ are subsets of ${\cal C}_2$.
\par
 Now, we trace the motion of an arbitrarily chosen EBs in exactly the same way
 as before except that when the tracing EB interacts with the EB that leads to
 ${\cal C}_1$ on its way, then we turn to trace the motion of resultant right
 EB. Now, we write ${\cal C}_2$ as the highest affinity EB under this new
 monitoring scheme in ${\mathbf A}^{N-3} (\alpha)$. Then, using the same
 argument as before, it is clear that all EBs in ${\mathbf A}^{N-3} (\alpha)$
 but ${\cal C}_1$ are subsets of ${\cal C}_2$. Besides, ${\cal C}_2 \subset
 {\cal C}_1$. Inductively, by tracing the motion of EBs in a similar way that
 leads to ${\cal C}_1$ and ${\cal C}_2$, we conclude that the EBs in
 ${\mathbf A}^{N-3} (\alpha)$ interact elastically. Hence, by
 Lemma~\ref{Lemma:Omega_i}, this theorem is proved.
\hfill$\Box$
\par\medskip\indent
 With the help of Theorem~\ref{Thrm:Time_Affinity_Rule}, the $n$ary density
 classification problem using two CAs can be easily executed.
\begin{define}
 We express a configuration as a collection of EBs. For any three consecutive
 EBs ${\cal C}_1 {\cal C}_2 {\cal C}_3$, we define the propagation rule
 ${\mathbf P}$ as follows: The states of sites in ${\cal C}_2$ remain unchanged
 if none of the three blocks ${\cal C}_1$, ${\cal C}_2$ and ${\cal C}_3$ equals
 ${\cal B}_i$ for all $0\leq i < n$. Otherwise, we set the state of sites in
 ${\cal C}_2$ to $i$ where $i$ is the state of the minimum affinity EB among
 ${\cal C}_1$, ${\cal C}_2$ and ${\cal C}_3$. \label{Def:Propagation_Rule}
\end{define}
\par\medskip\indent
 Since we can only find at most one type of ${\cal B}_i$ ($0\leq i < n$) in
 ${\mathbf A}^{N-3} (\alpha)$, repeatedly applying the propagation rule
 ${\mathbf P}$ to ${\mathbf A}^{N-3} (\alpha)$ results in propagating that
 particular EB ${\cal B}_i$. In addition, similar to the affinity rule, it is
 straight-forward to show that ${\mathbf P}$ is a CA rule with a radius
 $2n\!-\!1$. Besides, we have
\begin{theorem}
 If $\alpha$ is a configuration whose EB only interacts elastically, then
 applying ${\mathbf P}$ $\left\lceil \left\lceil \frac{N-1}{2} \right\rceil /
 2 \right\rceil$ times to $\alpha$ will result in having all $i$s in the
 configuration if and only if $\alpha\in\Omega_i$.
 \label{Thrm:Time_Propagation_Rule}
\end{theorem}
\par\medskip\noindent
{\em Proof:} Lemma~\ref{Lemma:Omega_i} tells us that $\alpha$ contains the EB
 ${\cal B}_i$ if and only if $\alpha\in\Omega_i$ for $0\leq i < n$. That is,
 we can find at most one type of EB ${\cal B}_i$ for $0\leq i < n$ in $\alpha$.
 Clearly, the worst case occurs when $\alpha$ contains exact one such
 ${\cal B}_i$. Besides, all other EBs in $\alpha$ are of length two. In this
 case, it is clear that applying ${\mathbf P}$ $\left\lceil \left\lceil
 \frac{N-1}{2} \right\rceil / 2 \right\rceil$ times result in converting all
 sites to state $i$. Hence it is proved.
\hfill$\Box$
\par\medskip\indent
 In summary, we show that it is impossible to solve the $n$ary density
 classification problem in any dimension using one CA. Nevertheless, we can
 solve this problem by first applying the affinity rule ${\mathbf A}$ (with a
 radius between $n (2^{n-1}\!-\!1)(2^n\!-\!3)$ and $(2n\!-\!1) (2^{n-1}\!-\!1)
 (2^n\!-\!3)$ ) $N\!-\!3$ times and then followed by the propagation rule
 ${\mathbf P}$ (with a radius of $2n\!-\!1$) $\left\lceil \left\lceil
 \frac{N-1}{2} \right\rceil / 2 \right\rceil$ times. Since the run time of the
 two CA rules scales as $\mbox{O}(N)$, so apart from a constant speed up, our
 two CA $n$ary density classification rules are optimal. In other words, the
 dynamics of our two CA rules gives rise to $n$ stable fixed points each with
 configuration $i^N$. Moreover, we find a large number of unstable fixed points
 whenever there is more than one most frequently occurring state in a
 configuration. Finally, we remark that these CA rules are not unique and other
 equally good method exists \cite{Siu}.
\par
 It is instructive to extend the $n$ary density classification problem to
 rational function classification similar to that reported in Ref.~\cite{Chau}
 as well as to higher dimensional lattices. We plan to report these results in
 future.
\acknowledgments
 We would like to thank Jessica~H.~Y. Chan for her valuable discussions and
 P.~M. Hui for pushing us to investigate this problem.

\end{multicols}
\end{document}